\documentclass[12pt,preprint]{aastex}

\usepackage{graphicx}

\begin{document}

\title{Detailed Atmosphere Model Fits to Disk-Dominated ULX Spectra}

\author
{Yawei Hui, Julian H. Krolik} \affil{Department of Physics and
Astronomy, Johns Hopkins University, Baltimore, MD 21218;
\\ ywhui@pha.jhu.edu, jhk@pha.jhu.edu}

\begin{abstract}
We have chosen 6 Ultra-Luminous X-ray sources from the {\it XMM-Newton}
archive whose spectra have high signal-to-noise and can be fitted solely
with a disk model without requiring any power-law component. To estimate
systematic errors in the inferred parameters, we fit every spectrum to two
different disk models, one based on local blackbody emission (KERRBB) and one
based on detailed atmosphere modelling (BHSPEC). Both incorporate full
general relativistic treatment of the disk surface brightness profile,
photon Doppler shifts, and photon trajectories. We found in every case that
they give almost identical fits and similar acceptable parameters. The
best-fit value of the most interesting parameter, the mass of the central
object, is between 23 and 73~M$_\sun$ in 5 of the 6 examples.  In every
case, the best-fit inclination angle and mass are correlated, in the sense
that large mass corresponds to high inclination.  Even after allowing for
this degeneracy, we find that, with $\gtrsim 99.9\%$ formal statistical confidence,
3 of the 6 objects have mass $\gtrsim 25~M_\sun$; for the other 3, these
data are consistent with a wide range of masses.  A mass
greater than several hundred $M_\sun$ is unlikely for the 3 best-constrained
objects.  These fits also suggest
comparatively rapid black hole spin in the 3 objects whose masses are relatively
well-determined, but our estimate of the spin is subject to significant
systematic error having to do with uncertainty in the underlying surface
brightness profile.
\end{abstract}


\section{Introduction\label{sec:intro}}

Nearly two decades since their first unveiling in the X-ray sky
\citep[][]{fabbiano1987},
Ultra-Luminous X-ray sources (ULXs) still remain puzzles. The X-ray
luminosities of these off-galaxy-center point
sources fall in the range $10^{39}$ -- $10^{41}$~erg~s$^{-1}$ in the
2 -- 10~keV band, at least an order of
magnitude larger than those of Galactic X-ray Binaries (GXB) and several
orders of magnitude smaller than those of Active Galactic Nuclei (AGN). If
ULXs are accreting black hole systems that observe the Eddington limit
($L_E=1.3\times10^{38}$~erg~s$^{-1}$), the central black holes must have
masses from a few tens to a few thousands of solar masses, i.e. they must be
intermediate mass black holes (IMBH). Some evidence has been found
in support of this view \citep[see][]{colbert1999,miller2002,miller2003}.
But it is also possible that ULXs' masses are smaller than the Eddington
limit suggests.  They might be stellar mass black holes whose radiation is
beamed in the observed direction \citep{king2001} or which are accreting at
super-Eddington rates \citep{begelman2002,begelman2006,stern2006}.  The only
way to distinguish between these scenarios is to measure the central masses
of ULXs as accurately as possible.

The most reliable way to measure the masses of the stars in a binary is to
observe their orbital motions. Unfortunately, this method does not work
in the context of ULXs. Because there are so far no clearly identified
spectral lines in ULXs' spectra, the common technique used in spectroscopic
binaries cannot be applied. Temporal variability is another approach. 
After comparing the Fourier power spectra of GXBs and AGNs, some researchers
(e.g.~\citet{mchardy2006}) have concluded that the characteristic timescales
of accreting black holes are proportional to their central objects' masses.
If ULXs are extragalactic analogs to the GXB
systems in our galaxy, similar correlations between quasi-periodic
oscillation (QPO) frequencies and mass may be used to estimate ULX
masses. These correlations have been established successfully for several
ULXs, for example, M82~X1 \citep{strohmayer2003} and NGC~5408~X1
\citep {strohmayer2007}. The discovery of the QPO in NGC~5408~X1 led to an
estimated ULX mass of 1500 -- 3500~M$_\sun$. However, these works treated
a limited number of ULXs and it is not guaranteed that QPOs can be found
in a large sample of ULXs.  Even if there is a QPO identified in the power
spectrum of a ULX, the mass inferred through this method is highly dependent
on the ``calibration standard'' and may be quite uncertain
\citep[see, for example][]{strohmayer2007}. 

Another method for inferring the central mass is spectral fitting
\citep{vierdayanti2006,winter2006,winter2007}. To fit a ULX spectrum,
various models and/or model combinations can be exploited. In practice,
ULXs' spectra are usually modelled by a combination of a thermal disk component
and a power-law \citep{miller2003,miller2004,roberts2004}). In this combination,
the thermal disk component covers the soft X-ray (0.3 -- 1 or 2 keV) range of the
spectrum while the power-law takes care of the hard tail (1 -- 10 keV and
above). Because the spectrum of soft X-ray radiation from the accretion disk
is dependent on the central mass, but there is no clear relation between the
hard tail of the spectrum and the central mass, we define our sample to
include exclusively those ULXs with only
thermal disk components in their spectra. In this way, we can
focus purely on the component of the system --- the thermal accretion disk
--- with the tightest connection to the central object's mass, as well as the other
key accretion parameters: spin, accretion rate and inclination angle.

There are many thermal disk models that have been used in this way.
In Xspec11, for instance, there are DISKBB, DISKPN, GRAD, KERRBB.
These models share more or less the same core --- the multi-color disk
(MCD) approximation \citep{mitsuda1984,makishima1986,shimura1995}.
In this approximation, one assumes that the disk emission is a sum of a
series of local blackbody spectra with effective temperatures defined by the
standard disk model but modified by a hardening factor.  The version
we employ here is KERRBB, which also incorporates general relativistic
effects.  Even though the
MCD approximation works well in the case of GXB \citep{davis2005}, in
this paper, we also fit the data to a model
in which a detailed stellar atmosphere is computed for each disk annulus.
Our reason to do so is that earlier work \citep{hui2005} has shown that
the spectral shape, especially around the peak energy, can be altered
by atomic features when the central mass is larger than typical stellar
masses ($>$~100~M$_\sun$).  This model (which also fully includes general
relativistic effects) has been compiled to an easy-to-use
table model in Xspec11 (BHSPEC: \citet{davis2006}); we modified
it and extended its parameter space to cover a wider range of
central masses (see \S~4 for details).  

In \S~2 we describe the sample selection criteria and in \S~3 the data
reduction procedure.  The disk model
BHSPEC is briefly discussed, and specific fitting results for each object
are presented in \S~4. For comparison, previous work on the objects in
our sample is briefly described in \S~5.  We discuss our results in \S~6
and draw conclusions in \S~7.

\section{Sample selection --- disk dominated ULXs}\label{sec:sample_data}

Our sample was drawn from the {\it XMM-Newton} public data archive.
We began with the object lists compiled by Winter et.~al. (see Table 6 in
\citet{winter2006} and Table 1 in \citet{winter2007}).  For their
sample, Winter et~al. chose objects observed for at least 10~ks and with
distance less than 8~Mpc.  These criteria guaranteed a minimum of 400 counts
for objects with $L_X>2 \times10^{38}$~erg~s$^{-1}$.  To focus on ULX
candidates, they also removed all objects with spectra distinctly different
(e.g., resembling supernova remnants) and all objects located at the
centers of their host galaxies (to eliminate AGN).  We chose those ULX
candidates with $>$~1000 counts in order to guarantee good signal-to-noise in
the reduced spectra. Then, in order to retain as much as possible of the
soft X-ray emission from the accretion disk, we excluded those objects with
neutral hydrogen column density in the line of sight larger than
$10^{22}$~cm$^{-2}$ as determined by Winter et~al. (see Table 4 in
\citet{winter2006} and Table 2 in \citet{winter2007}), except for
Circinus~X1 and X2, which have slightly higher values. There are 23 objects meeting
these criteria. 

We then used the latest version (7.0) of the {\it XMM-Newton} Science
Analysis System (SAS) and calibration (up to date as of Jan. 2007) to reprocess
the data and obtain the event files. After
applying proper filters (see \S~3 for details), source and
background spectra for each of these objects were
created and fitted by the ``disk+powlaw'' combination in Xspec11.
Specifically, we used a photoelectric absorption model (phabs) to
account for the opacity in the interstellar medium and the thermal disk model
BHSPEC (a detailed disk model discussed later in this paper) plus a
power-law component.  To decide which model component (disk or powlaw)
dominates the spectral output, we computed the integrated counts
from both components and defined the spectrum to be dominated by one or the
other when one component contributes more than 70\% of the total counts.
There are 11 powlaw-dominated spectra, 6 disk+powlaw spectra,
and only 6 objects (see Table~\ref{tb-diskULXs}) having
disk-dominated spectra. In other words, we can use the disk
model alone to fit those 6 disk-dominated spectra. In these objects, a non-zero
power-law contribution at best marginally improves the quality of the fit
and sometimes even harms it.

\section{Data reduction}\label{sec:reduction}
We used the latest version (7.0) of SAS and the up-to-date calibration
(Jan. 2007) to reprocess the Observation Data Files (ODF). Commands
``emchain'' (for EPIC-MOS) and ``epchain'' (for EPIC-PN) in SAS 7.0 were used
to get the new processed pipeline
products (PPS) which include the photon event files. Standard
data filter procedures were followed according to the instructions in the
{\it XMM-Newton} ABC guide. Good events were required to satisfy the
following conditions: 1) for the MOS
(both MOS1 and MOS2) detectors, the event pattern is in the 0 to 12 range
(single, double, triple, and quadruple pixel events) and the pulse height is
in the range of 0.2 -- 12 keV; 2) for the PN detector, the event pattern is in
the 0 to 4 range (single and double pixel events) and the pulse height is in
the range of 0.2 -- 15 keV. The \#XMMEA\_EM (for the MOS) filter and \#XMMEA\_EP
(for the PN) filter were also applied together with a ``FLAG==0'' filter to
kick out any bad or close-to-edge events captured by the CCDs. The light curve
for each observation was produced with the SAS command `evselect'' to decide a
proper time filtering
threshold in order to reduce the influence from high rate flaring. For
the MOS detectors, the good time intervals were selected by setting the
rate to less than 5~cts~s$^{-1}$ (sometimes a little bit higher); for the
PN detector, the threshold varies from 25 to 60~cts~s$^{-1}$.

After obtaining the filtered event files, source and background spectra were
extracted by using the SAS procedure ``especget''. For the sources, circular
regions with radii of 20
arcseconds ($\pm10$ according to the size of the sources and closeness to
other sources or detector edges) were applied in the spectra extraction. For the
background, a nearby circular region close to but without overlap on the
source's region was used, with radius double the size of its corresponding
source. SAS procedures ``rmfgen'' and ``arfgen'' were used to generate the
response matrix files (RMF) and ancillary response files (ARF). Finally, we
used the command in HEASOFT 6.1 --- ``grppha'' --- to regroup the spectra so that
there are at least 20 photon counts in every spectral bin.

\section{Model description and fitting results }\label{sec:model_fit}

\subsection{BHSPEC vs KERRBB}\label{subsec:model}
We fitted the selected spectra with both BHSPEC (a detailed atmosphere
calculation) and KERRBB (a MCD model).  For details
of KERRBB, see \citet{li2005} and the Xspec11 manual; the only significant
choice we made was to set the dilution factor to 1.7.  BHSPEC is
based on an atmosphere model of a standard $\alpha-$disk \citep{shakura1973}
where each ring is treated in hydrostatic equilibrium and energy balance.
The stress parameter $\alpha$ is chosen to be 0.01 so that all disk
annuli are optically thick.  The radiation intensity is computed as a
function of both frequency and position by means of a full radiation transfer
solution.  Continuum opacities due to free electrons, H, He, C, N, O, Ne, Mg, Si, S,
Ar, Ca, Fe and Ni, with the abundances given by \citet{anders1989} are
calculated on the basis of explicit ionization balance calculations and
statistical equilibrium of the most populated internal states.  Photon-electron
energy exchange by Comptonization is also included in the transfer solution.
General relativity effects in photon propagation are taken
into account: radiation is boosted and beamed, and its trajectories ``bent''
in the general relativistic potential before it
reaches infinity \citep{agol1997}.  For further details, see \citet{hui2005}.

In the work of \citet{davis2006}, disks with many black hole masses, spins,
accretion rates, and inclination angles were modelled, and the spectra obtained
were compiled into the Xspec11 table model BHSPEC. 
For this project, we extended the parameter space considered by Davis and
Hubeny to include disks around black holes with masses in the range of 10 --
10$^4$~M$_\sun$.  Because the internal consistency of disk
models becomes dubious when the luminosity approaches or exceeds the
Eddington luminosity, we included no models with $L/L_E > 1$.  When
we speak of lower bounds on the mass based on BHSPEC-fitting, we
therefore always mean that this bound is subject to the constraint of
being consistent with a disk model, i.e.,
keeping the luminosity sub-Eddington.  It should be borne in mind,
however, that some significant flexibility nonetheless remains at the low-mass
end of parameter space because altering the inclination angle or the
black hole rotation rate could keep the luminosity in our direction at
the level required by the measurements even while the luminosity in other
directions is diminished in order to satisfy the sub-Eddington constraint.
We did not so restrict KERRBB, so any super-Eddington fits it produces must
be regarded as extremely suspect because the underlying model would then
be inappropriate.  For technical reasons, we limited the range of spins
considered by BHSPEC to $0 \leq a/M \leq 0.997$ and by KERRBB to
$-0.999 \leq a/M \leq 0.999$.

\subsection{Fitting results}\label{subsec:fit_results}
The fitting results for the objects in our sample are shown in
Table~\ref{tb-fitResults}. Each object is given two rows. The first row, with
the object's name, describes the fit with BHSPEC; the second row shows
the result from KERRBB. In BHSPEC, the normalization is fixed at a
value corresponding to the best estimate for that object's distance
(tabulated in Table~\ref{tb-diskULXs}). All other parameters, i.e.,
the equivalent hydrogen column density of photoelectric absorption, the black
hole mass and spin, the
accretion rate, and the disk viewing angle are set free to vary. In KERRBB,
the parameter ``distance'' is fixed to the same value as in BHSPEC for each
object and the normalization is set to 1.0. KERRBB also offers several other
parameters. We freeze ``eta'' to zero (the zero torque inner boundary condition),
``hd'' to 1.7 (common diluted black body spectrum), ``rflag'' to $-$1.0
(self-irradiation is off), and ``lflag'' to 1.0 (limb-darkening is on).

The best-fit values and their uncertainties are shown in
Table~\ref{tb-fitResults}. All are statistically acceptable, and in each case the
BHSPEC least $\chi^2$ is slightly smaller than the KERRBB value. All
one-parameter uncertainties were computed with
$\Delta\chi^2=2.706$, equivalent to 90\% confidence for a single parameter.
For the BHSPEC fits, besides the best values, we also calculated the ratio
of the observed luminosity ($E=0.3$ -- $10$ keV, except for Circinus~X2
MOS1, in which case the integration is limited in $E=0.55$ -- $10$ keV)
to the computed total intrinsic luminosity ($E=0$ -- $\infty$ keV) emitted
in our direction; these are shown in the last column
of Table~\ref{tb-fitResults}.   The fact that all these ratios are close
to unity demonstrates that the {\it XMM-Newton} energy band contains
nearly all the light our models suggest is being radiated.

In the case of KERRBB, the spectral fitting gives the accretion rate in the
form of $\dot{M}$ instead of the
dimensionless $L/L_E$. We used the following formula to translate between
those two definitions, assuming that the radiative efficiency, $\eta$, is
the standard function of $a/M$ \citep{novikov1973}.
\begin{displaymath}
\frac{L}{L_E}=\frac{\eta \dot{M} c^2}{1.3\times10^{38}M}.
\end{displaymath}
Here $c$ is the speed of light, $\dot{M}$ is the accretion rate in units of
gm~s$^{-1}$ and $M$ is the accretor's mass in units of M$_\sun$. Because the
total luminosity of an object is fixed when the distance is known, the ratio 
$L/L_E$ is actually inversely proportional to the object's mass. For this
reason, only the uncertainty in the estimated mass is taken into account
in the quoted uncertainty for $L/L_E$.

\begin{figure}
\mbox{
\includegraphics[angle=270,scale=0.35]{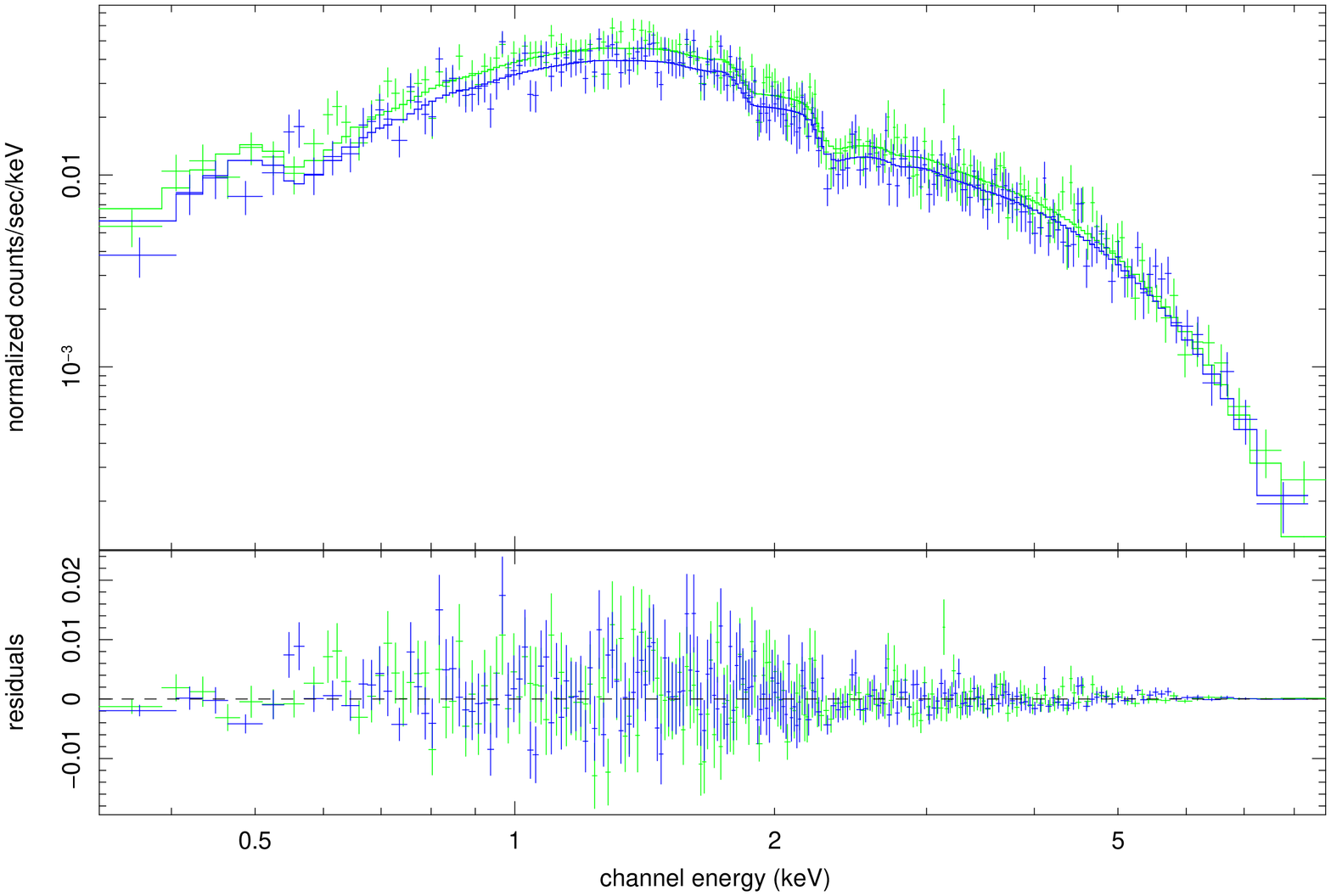}
\includegraphics[angle=270,scale=0.35]{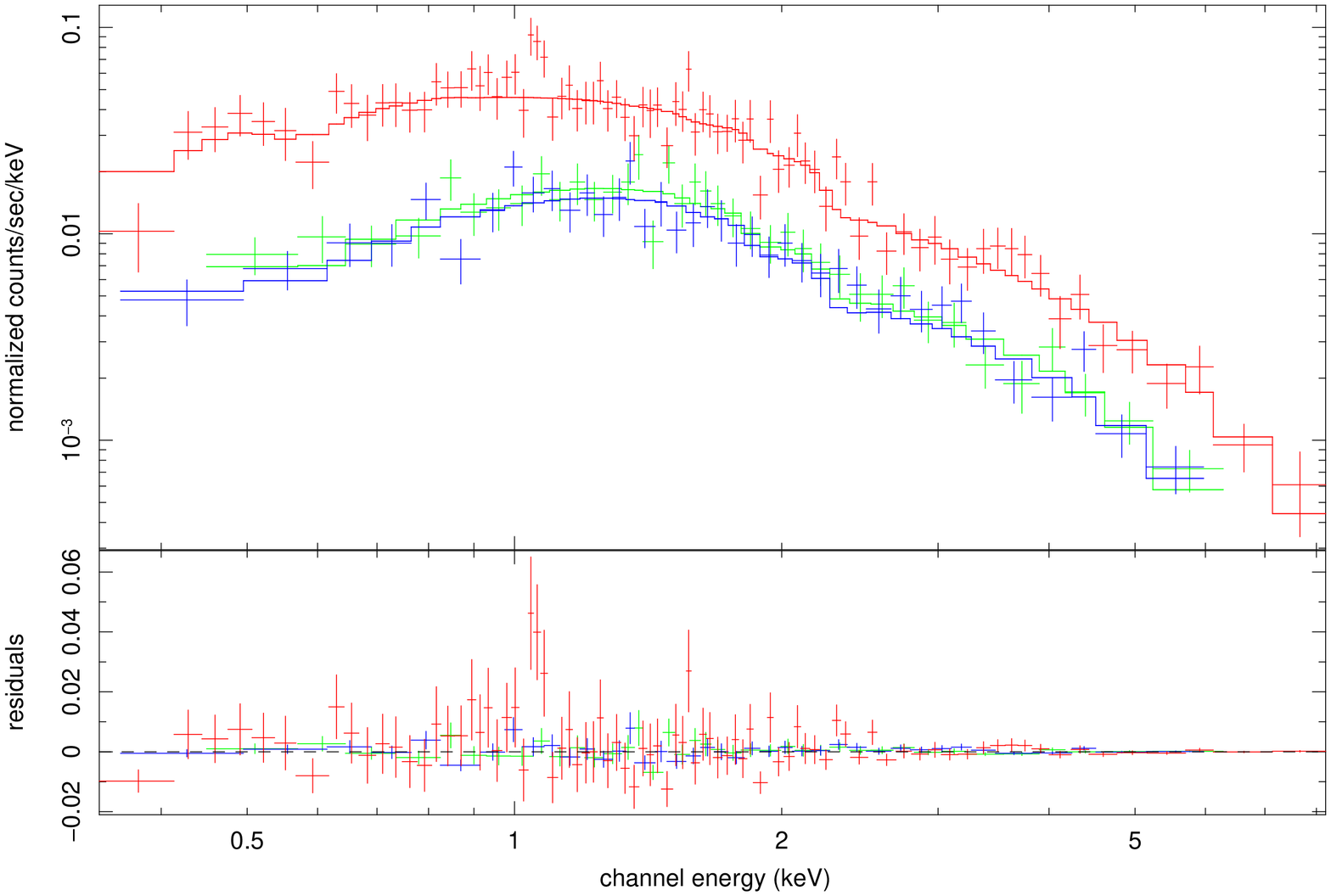}
}\par
\mbox{
\includegraphics[angle=270,scale=0.35]{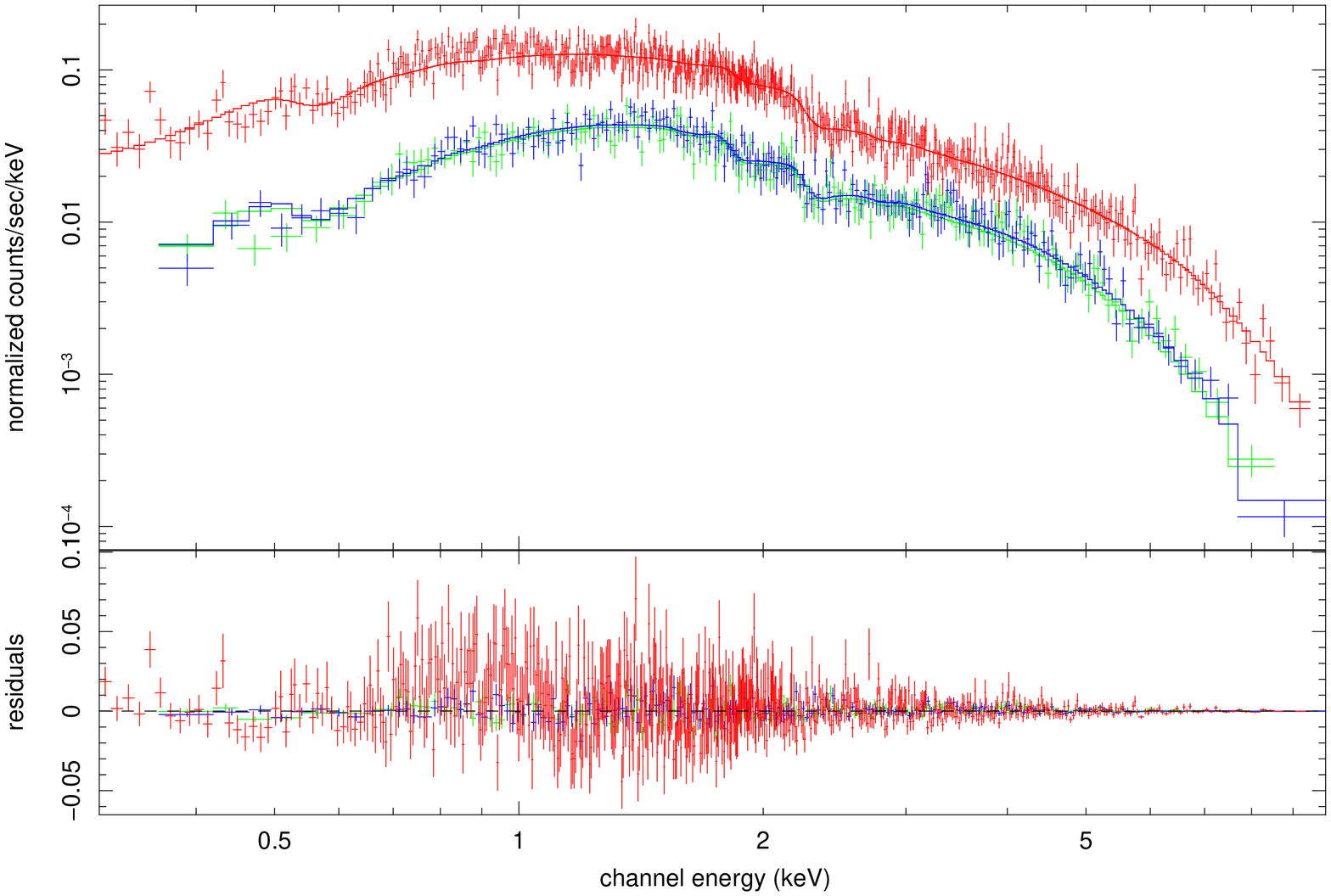}
\includegraphics[angle=270,scale=0.35]{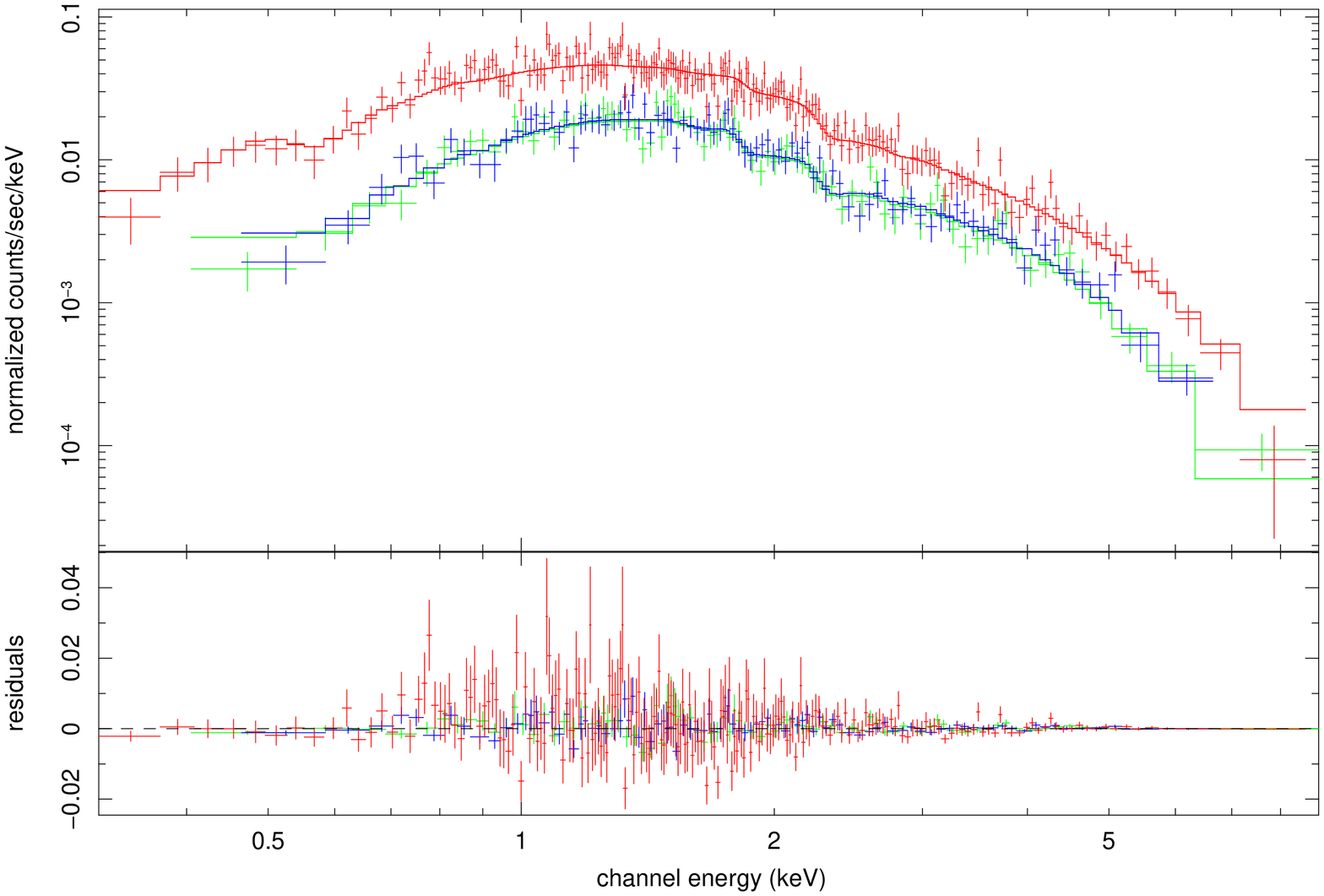}
}\par
\mbox{
\includegraphics[angle=270,scale=0.35]{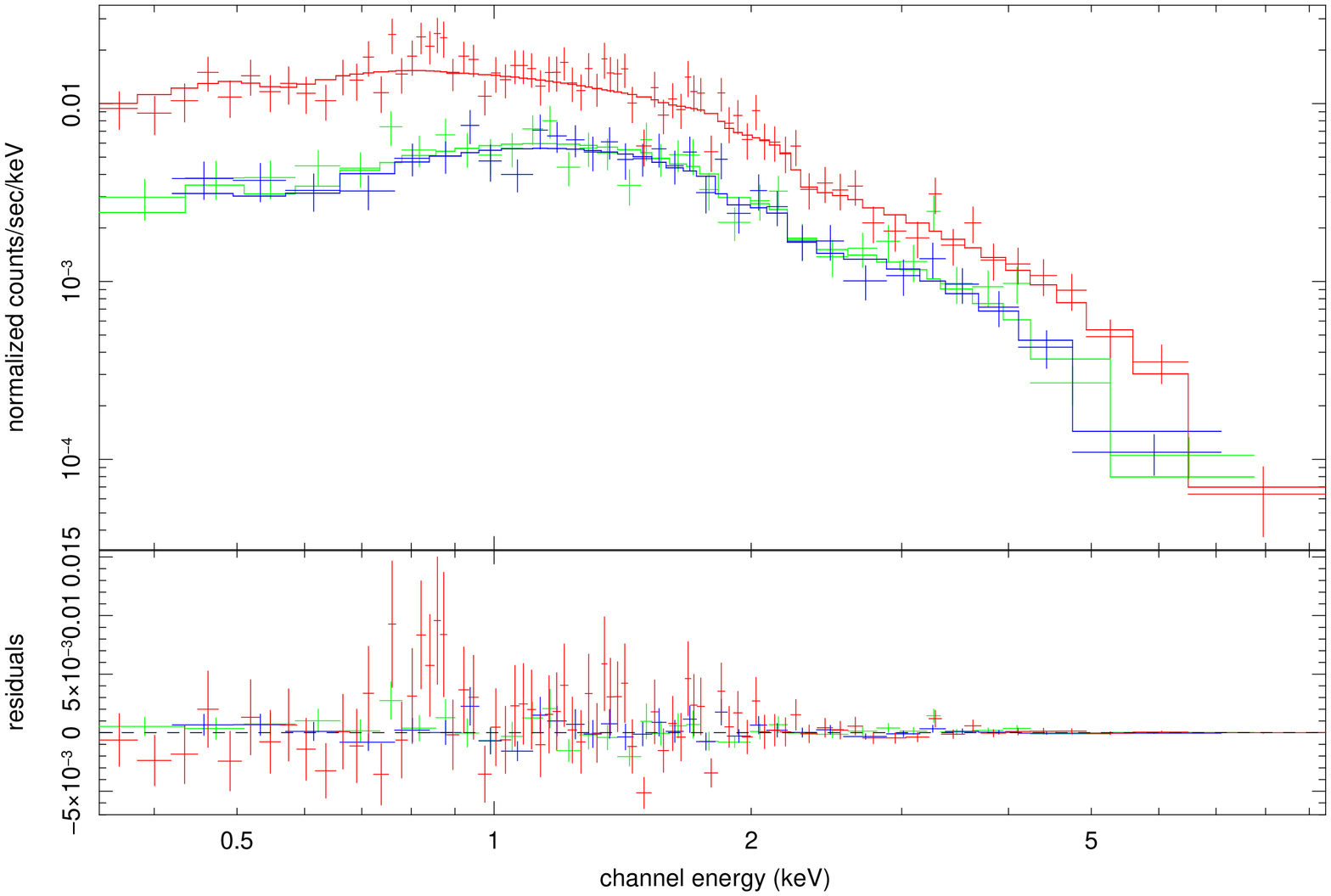}
\includegraphics[angle=270,scale=0.35]{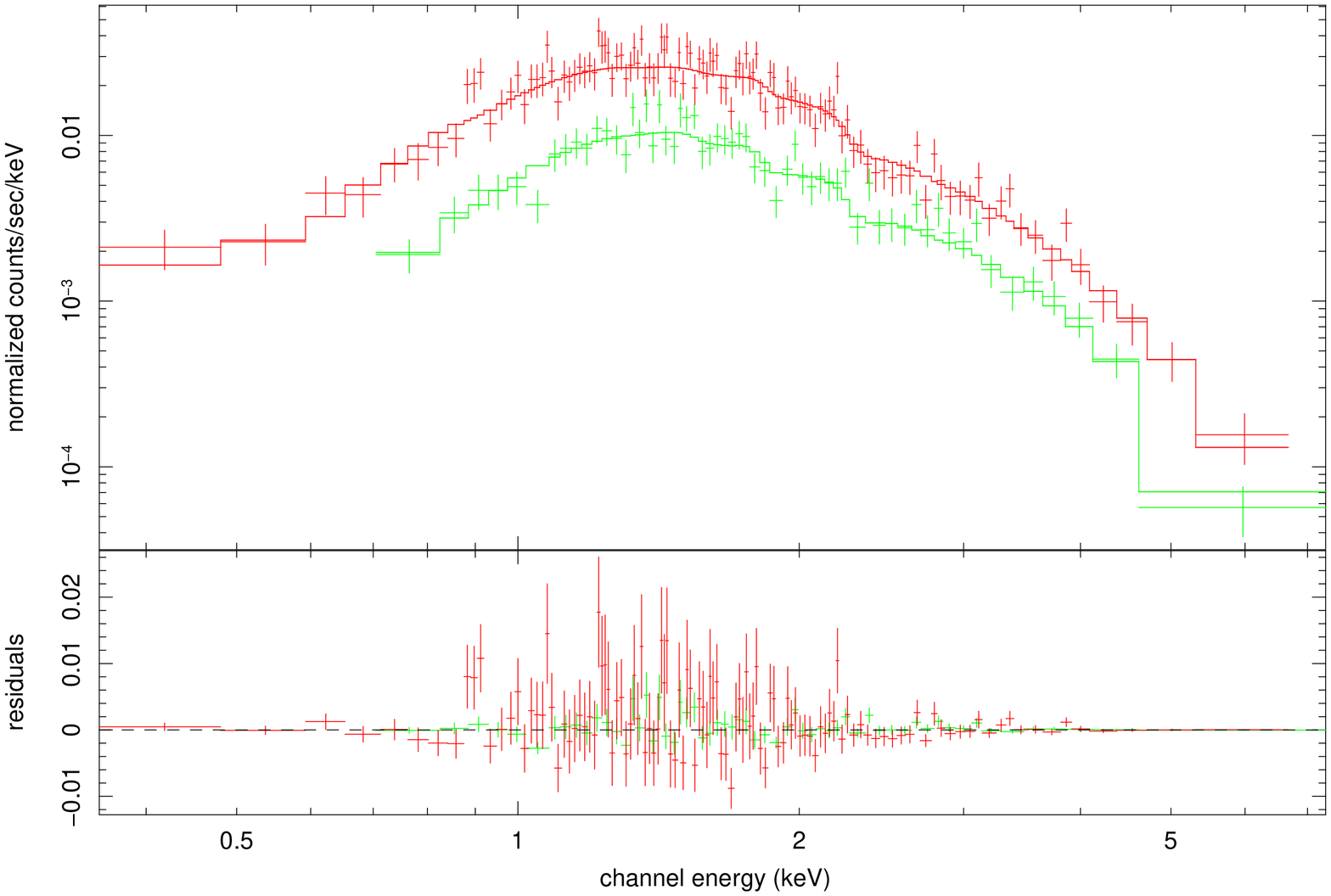}
}
\caption{Observed spectra and best-fit models (only BHSPEC shown). Panels from top
to bottom: upper left --- M81 X1, upper right --- M101 X1, middle left ---
NGC~253 X1, middle right --- NGC~253 X3, lower left --- NGC~253 X4, lower
right --- Circinus X2. In each panel, the green, blue and red colors represent the data from
MOS1, MOS2 and PN, respectively. \label{fig-spectra-fits}}
\end{figure}

In the following subsections, we discuss each object individually. The
spectra and best-fit models (BHSPEC only) are shown in the multipanel
Figure~\ref{fig-spectra-fits}.  The KERRBB fits are so nearly identical
to the BHSPEC fits that figures displaying them would be visually
indistinguishable from each other.

\subsubsection{M81~X1}\label{subsubsec:m81x1}
We used a 120~ks {\it XMM-Newton} observation (0200980101) of Holmberg IX
(made in September 2004) for M81~X1, which appears in the same field.
We extracted spectra from the event files generated for the MOS1 and MOS2
cameras and fitted them with BHSPEC and KERRBB. The mass of the
black hole is inferred to be 67 -- 85~M$_\sun$ from BHSPEC, while KERRBB gives
a partially overlapping, but rather wider possible range, 33 -- 74~M$_\sun$.
For the accretion rate, both
BHSPEC and KERRBB indicate values close to the Eddington limit, which
makes the applicability of both models suspect because both assume thin
disks.  Both models also give
similar values for the hydrogen column density (1.5 -- 2.0
$\times10^{21}$ cm$^{-2}$), the black hole spin (close to the maximum Kerr
value) and inclination angle (50$\degr$ - 70$\degr$). 

\subsubsection{M101~X2}\label{subsubsec:m101x2}
M101 was observed (0104260101) in June 2002 for a duration of 43 ks. 
We extracted spectra for the MOS1, MOS2 and PN cameras. The BHSPEC fit suggests
that the black hole in M101~X2 has a mass 30 -- 178~M$_\sun$ and the spin
of the black hole is close to the maximum Kerr value. The accretion rate may
be anywhere from the Eddington limit down to a moderate level
($L/L_E \simeq 0.3$). The viewing angle is inferred to be
42$\degr$ -- 78$\degr$.  Using KERRBB indicates a lower hydrogen column density
and larger uncertainty ranges for the mass and accretion
rate even though the best-fit values are not far from those found using BHSPEC.
The spin and the viewing angle in the KERRBB fit are totally unbounded.

\subsubsection{NGC~253~X1, X3 and X4}\label{subsubsec:ngc253}
The observation (0152020101) of NGC~253 was made in June~2003 with a
duration of 140~ks. The spectra were drawn from the event files of the MOS1,
MOS2 and PN cameras.

For NGC~253~X1, both BHSPEC and KERRBB indicate similar hydrogen
column densities. For the black hole mass, BHSPEC infers a 90\% confidence
range of 50 -- 81~M$_\sun$, while KERRBB gives a very similar, but
slightly wider range, 54 -- 95~M$_\sun$. In both models, the spin
is quite large and very close to the maximum Kerr value. The
Eddington-normalized luminosity inferred by BHSPEC is $0.35 < L/L_E <
0.56$, while KERRBB suggests a lower range, $0.18 < L/L_E < 0.31$. The
viewing angle is well constrained by
both models in the range $64\degr$ - $83\degr$. 

In the case of NGC~253~X3, the uncertainty in the black hole mass derived
from both BHSPEC and KERRBB is somewhat greater, from 25 to 80~M$_\sun$ for
BHSPEC, from 32 to 176~M$_\sun$ for KERRBB.  In terms of $L/L_E$, the range
preferred by BHSPEC ($L/L_E$ from 0.13 to 0.39) overlaps that of
KERRBB ($L/L_E$ from 0.03 to 0.18), but for the most part suggests higher
values.  Both models give their best fits when the black hole spins rapidly. The
viewing angle is large, estimated to be between $44\degr$ and $85\degr$. 

The inferred black hole mass from BHSPEC for NGC~253~X4 lies in the range
10 -- 91~M$_\sun$; KERRBB gives a similarly large
uncertainty, from 5 -- 77~M$_\sun$.  For the normalized luminosity, the
two models have different best-fit values ($L/L_E= 0.14$ for
BHSPEC and  0.05 for KERRBB), but the ranges permitted within the errors
overlap to a considerable extent: 0.06 -- 0.24 (BHSPEC) and 0.02 -- 0.35
(KERRBB). BHSPEC suggests rapid black hole spin with a large viewing angle
(43$\degr$ -- 88$\degr$), while KERRBB puts no 
constraint on either of these two parameters.

\subsubsection{Circinus~X2}\label{subsubsec:cirx2}
Circinus~X2 was observed (0111240101) in August~2001 for a duration of 110~ks.
We extracted its spectra for both the MOS1 and PN cameras. 
Both BHSPEC and KERRBB failed to constrain almost all parameters except the
hydrogen column density (5.0 -- 5.9 $\times10^{21}$ cm$^{-2}$).
Unlike the other objects in our sample, the best-fit values of the mass
inferred by BHSPEC and KERRBB (i.e., 340~M$_\sun$ vs. 13~M$_\sun$) are
quite different, while the best-fit values of $\cos i$ are also very different
(0.07 vs 1.00).  This apparent conflict will be addressed later in the
discussion when we investigate the correlation between inferred
mass and inclination angle.

\section{Comparison to previous work}\label{sec:previouswork}

In a thorough study of M81~X1 (there named M81~X6) based on a May~2000
observation by {\it Chandra}, \citet{swartz2003} (hereafter, S2003)
showed that the
best-fit model for the reduced spectrum is an absorbed disk blackbody. This
disk domination in the spectrum is confirmed in our study. Swartz et.~al.
found the hydrogen column density (in units of $10^{21}$ cm$^{-1}$)
to be $2.17\pm0.10$, which is close to our fitting results,
$1.91^{+0.12}_{-0.11}$ for BHSPEC and $1.61^{+0.11}_{-0.11}$ for KERRBB. The
black hole mass they estimated was 18~M$_\sun$, but they did so
assuming the black hole is non-rotating. According to our
fitting results (for both BHSPEC and KERRBB), the spin of the black
hole in M81~X1 has close to the maximum Kerr value. These results are
compatible with our estimate that the mass is 33 -- 85~M$_\sun$ because the
(Boyer-Lindquist) radial coordinate of the innermost stable circular orbit
(ISCO) for a maximal Kerr BH is $\simeq \frac{1}{6}$ that of a Schwarzschild
BH of the same mass.

In S2003 and a previous {\it ASCA} study of M81~X1
\citep[][]{makishima2000}, the total luminosities were reported to be
equal to or even exceeding the Eddington limits of the inferred black hole
masses. To avoid this violation of conventional expectations, a bigger
black hole mass (to give a higher Eddington limit) with a larger spin
(to make a smaller ISCO) was suggested by
some authors.  As shown by our results, however, if one is constrained by
fitting the spectrum, there is a limit to how much the mass and spin can
be increased, and neither our atmosphere-based spectral model nor
the multi-color disk model is consistent with substantially sub-Eddington
behavior.

\citet{jenkins2004} studied the same {\it XMM-Newton} observation of the
galaxy M101 as we did, but named M101~X2 M101~XMM-1.  They fitted the spectrum
with a single disk component (DISKBB in Xspec) with satisfactory statistics
and obtained an inner disk temperature ($T_{in}=1.33$~keV). Due
to the limited predictive power of DISKBB, no further conclusions were
reached in their work.

\section{Discussion}\label{sec:discussion}

   Before dealing with our principal concern, the inferred masses of these
objects, we first discuss a more technical point: the surprising 
lack of difference between BHSPEC
and KERRBB when fitting the spectra. As described in \S~\ref{subsec:model},
BHSPEC is a much more sophisticated model than KERRBB and one might
expect that the atomic features it can predict might lead to interestingly
different fitting results.  The fundamental reason why this does not
happen is that, at the color temperatures of the objects in our sample,
even Fe is fully-stripped.   It is worth elaborating briefly on this
point because this constraint enters in a somewhat indirect fashion.
To zeroth order, the temperature of the disk scales as $L^{1/4}M^{-1/2}$.
If there were no relativistic Doppler shifts, the color temperature and
luminosity would then suffice to determine the mass.  However, disks
around black holes do offer large Doppler factors, opening a wider
parameter space in which to search for acceptable models.  For these
objects, we find that in part of this parameter space, the mass is
high enough ($\gtrsim 100$~M$_\sun$) that the temperature in the
disk drops to the point where the unstripped fraction of the heaviest
abundant elements can produce an interesting level of opacity
in atomic features \citep{davis2006,hui2005}.  However, the very fact
that the fluid-frame temperature is this low means that, in order
for the spectrum generated to fit the data, it must be Doppler
boosted, which always implies large inclination angle, and can be
enhanced by rapid black hole spin.  However, the non-uniformity of
the boost around the disk surface also entails strong smearing
of sharp features.  The result is that atomic features are never
apparent in any of our fits, even in the extreme case of Circinus~X2,
in which the best-fit mass is 340~M$_\sun$.  Moreover, because the
unstripped ion fractions are so small at thermodynamic temperatures close
to the observed color temperatures, this conclusion is only
very weakly dependent upon elemental abundances.  There is however,
one possible exception to these arguments: in the presence of some
atomic opacity, the error entailed in KERRBB by assuming a Planckian output
spectrum with a fixed dilution factor may be larger than in a
case in which there is truly zero atomic opacity.

\begin{figure}
\includegraphics[scale=0.7]{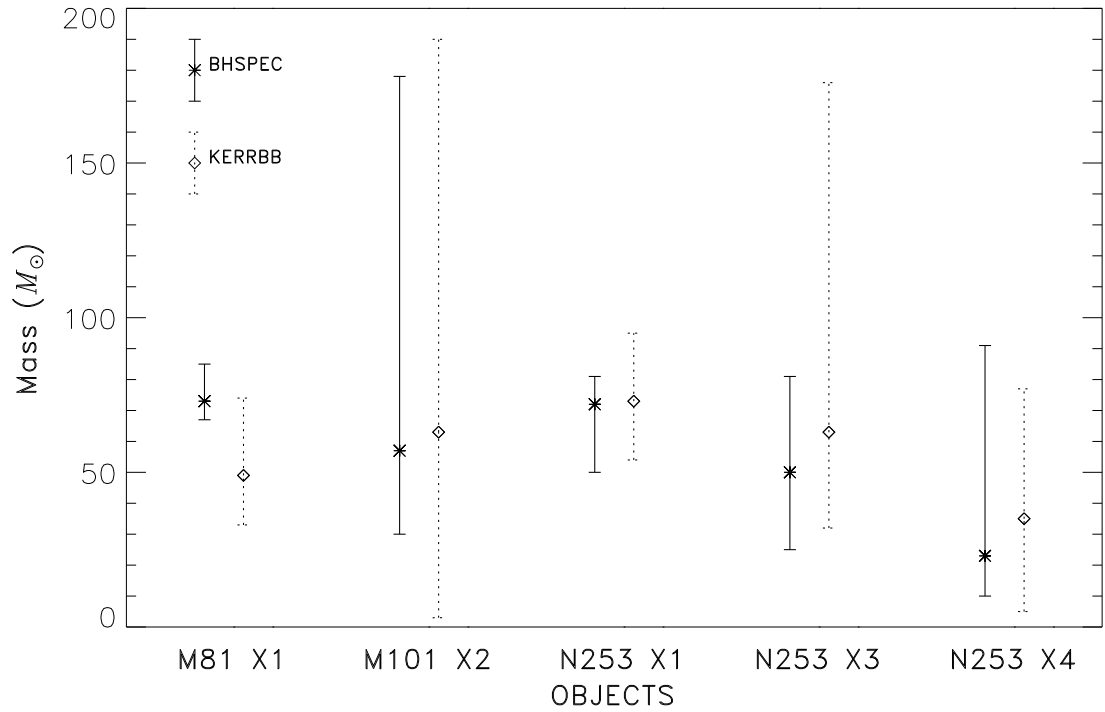}
\caption{The distribution of black hole masses in our sample (except
Circinus X2). Masses derived using BHSPEC are shown by stars, those
inferred on the basis of KERRBB are shown with open diamonds. All
uncertainties were computed with $\Delta\chi^2=2.706$, equivalent to
$90\%$ confidence for a single parameter.\label{fig-mass_obj}}
\end{figure}

We now turn to the main goal of this project: our attempt to use
spectral fitting of thermal disk spectra to constrain the masses of
ULXs.  In Figure~\ref{fig-mass_obj}, we plot the best-fit values and
uncertainties of the black hole masses for all our objects except Circinus~X2.
As that figure shows,
the best-fit values of the black hole masses span a surprisingly
small range, between 20 and 80~M$_\sun$.  In fact, the error bars are large
enough that the fits are consistent with all five having the same mass,
$\simeq 75 \pm 10$~M$_\sun$.

Moreover, this range appears to be quite different from that seen
in Galactic black hole binaries.  As reviewed by \citet{mcclintock2004},
of all 17 GXBs with measured black hole masses, there are none in which
the mass is likely to be smaller than 3~M$_\sun$ or larger than
18~M$_\sun$.

However, before we can conclude that these objects have a distinctly different
mass distribution than the Galactic black holes, it is important to look
more closely at the dominant source of uncertainty in the mass: the
degeneracy between $M$ and cos~$i$. \citet{sun1989} pointed out that
these two quantities are related almost inversely. In their disk fitting
analysis of quasars and Seyfert galaxies, they found
\begin{displaymath}
\mathrm{log}(M)\propto -b\cos i,
\end{displaymath}
where $b$ is a parameter that varies between 0.6 (for $a/M = 0$)
and $\simeq 1.2$ (for $a/M = 0.998$).  The origin of this relationship
lies in the fact that higher inclination produces greater blue-shifting
of light from matter orbiting toward us, thus compensating for the
diminution of temperature that generally occurs with larger central
mass (and therefore radiating area).  The coefficient is a function
of spin because the orbital speed near and outside the ISCO increases
with more rapid rotation.

\begin{figure}
\mbox{
\includegraphics[scale=0.35]{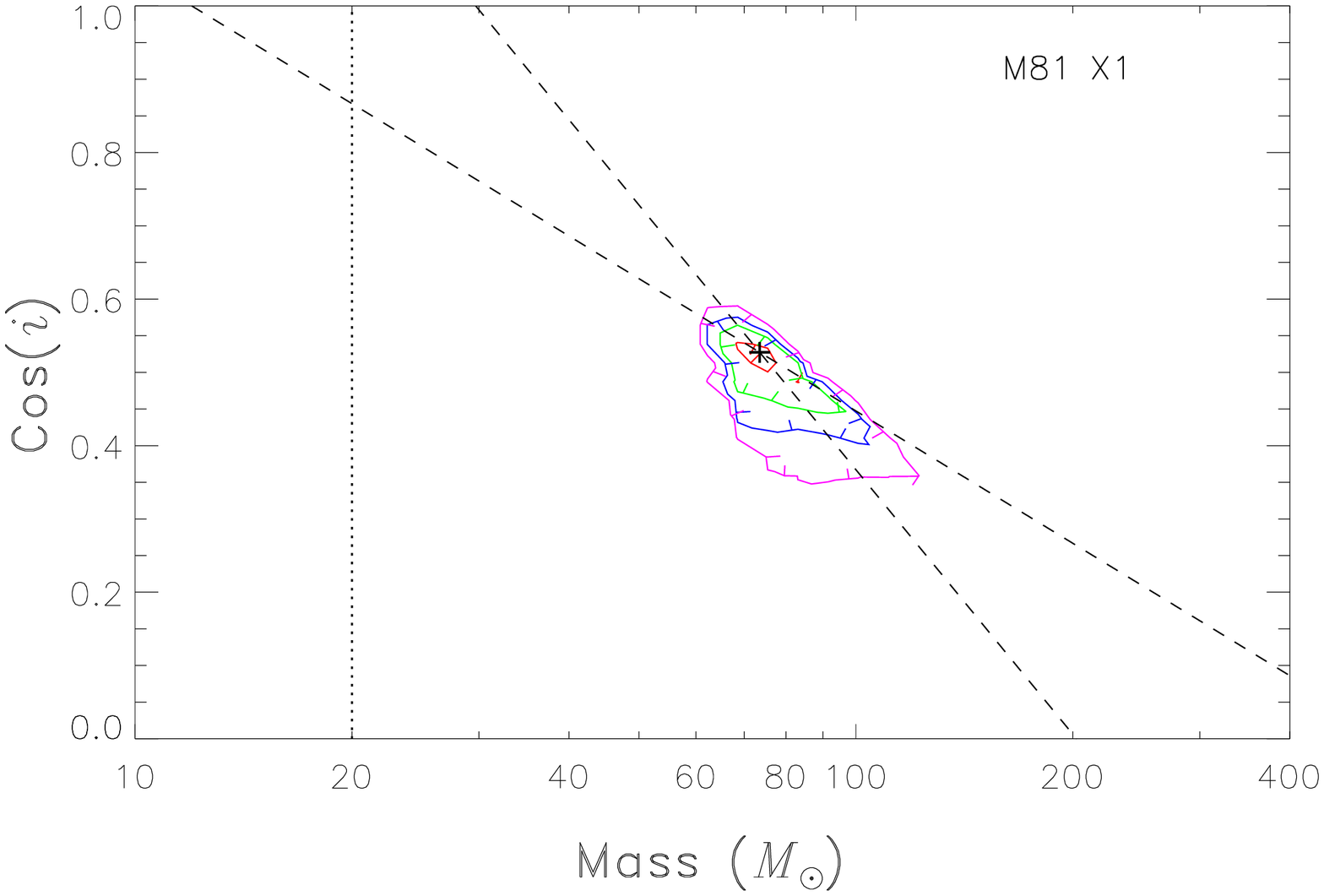}
\includegraphics[scale=0.35]{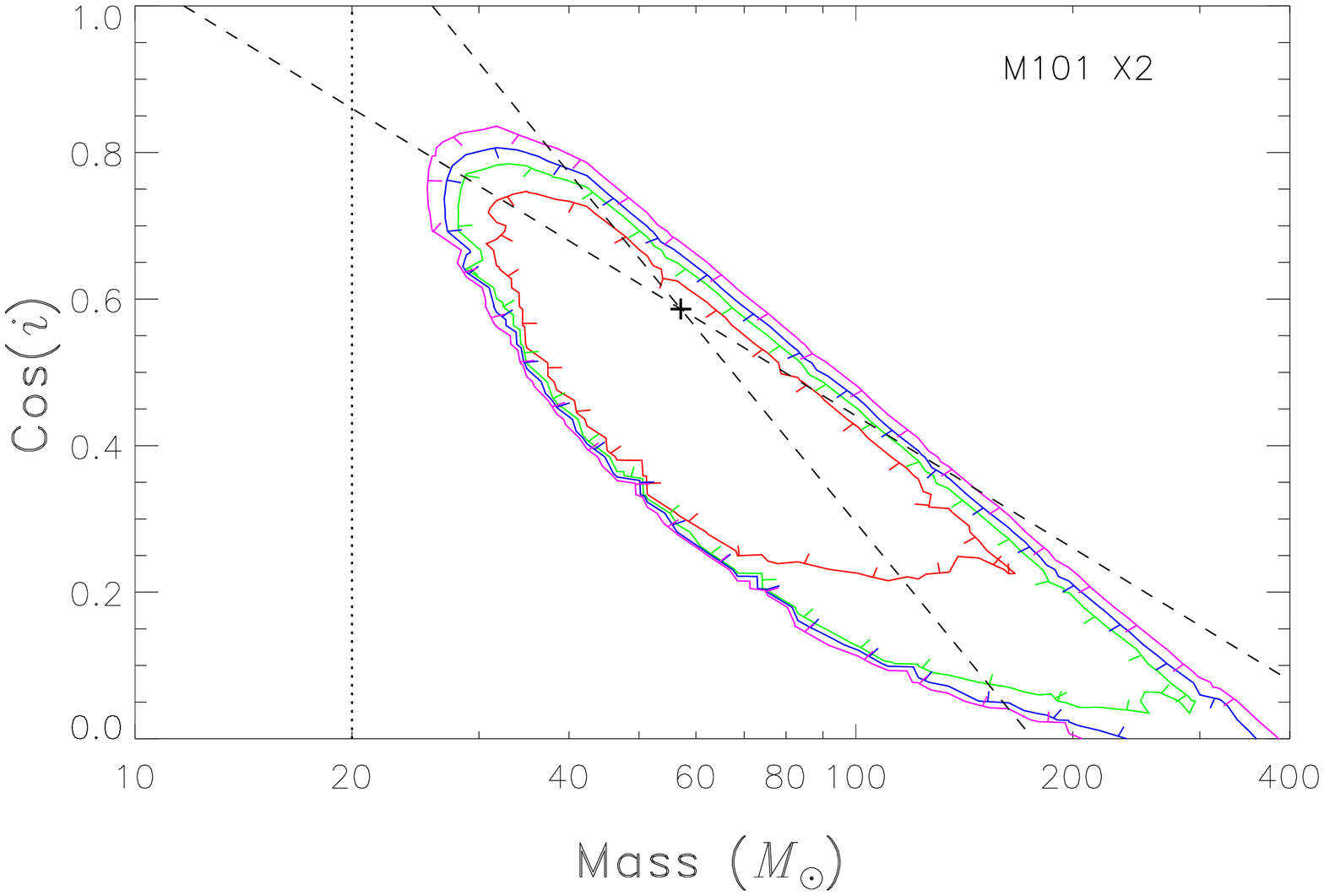}
}\par
\mbox{
\includegraphics[scale=0.35]{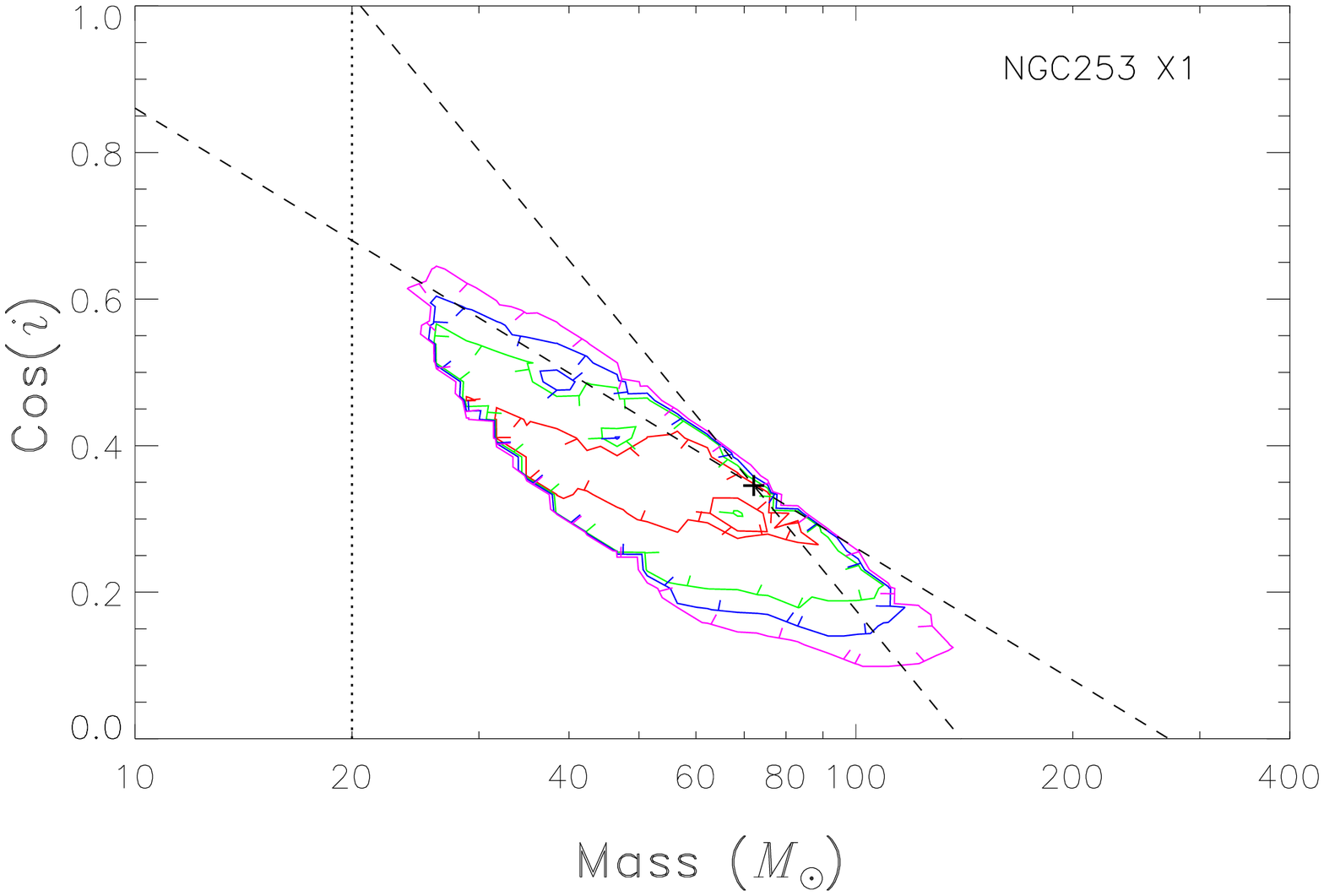}
\includegraphics[scale=0.35]{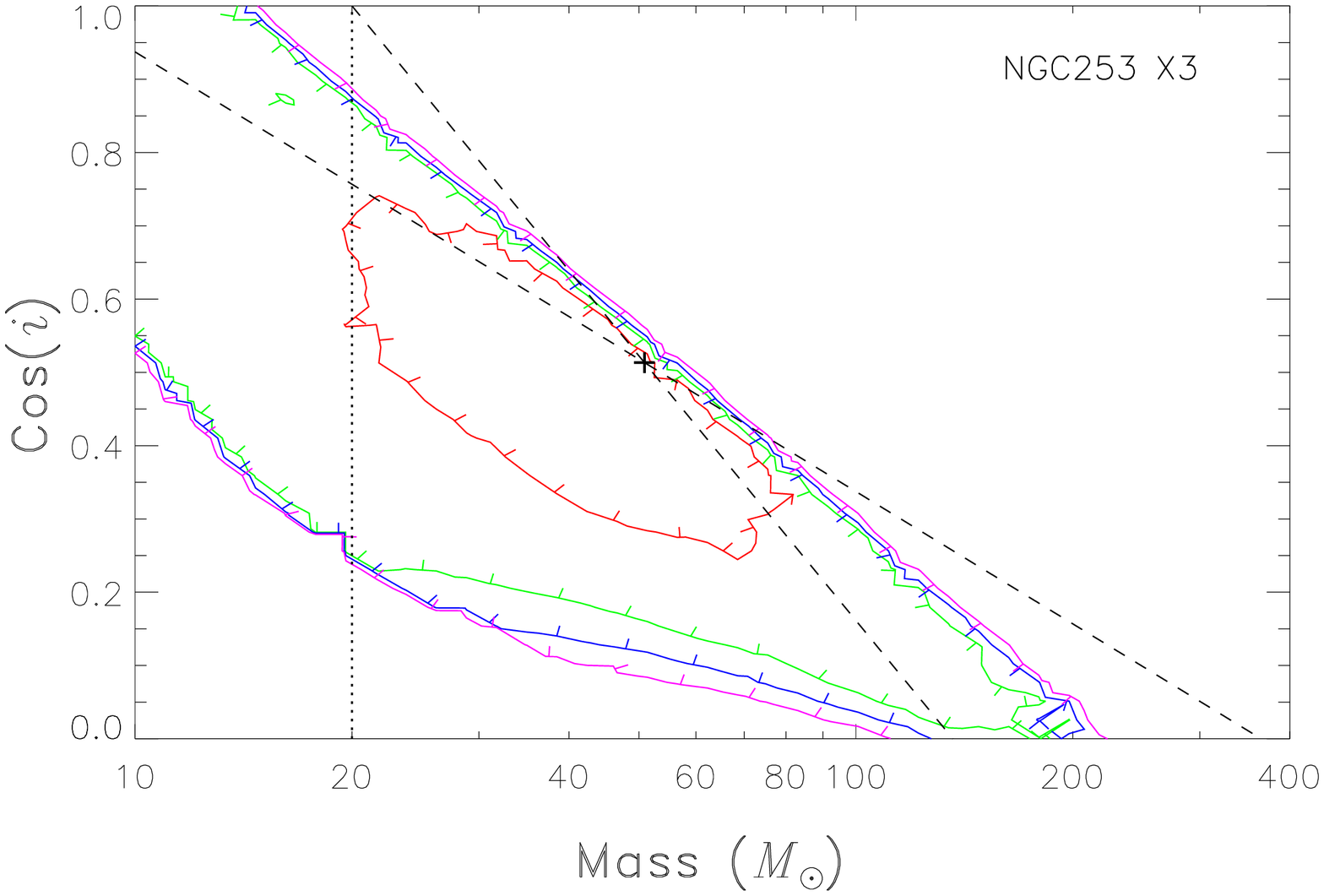}
}\par
\mbox{
\includegraphics[scale=0.35]{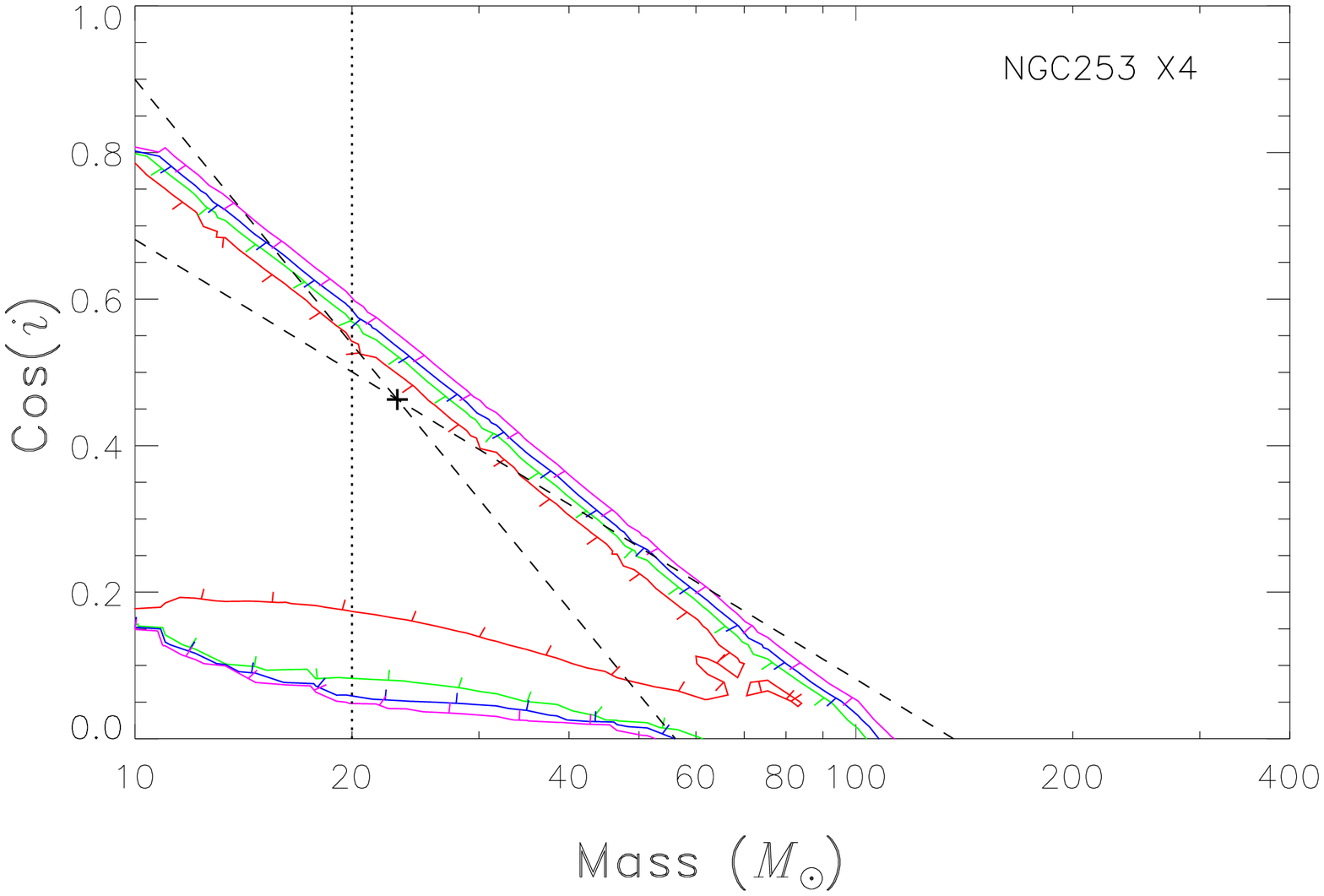}
\includegraphics[scale=0.35]{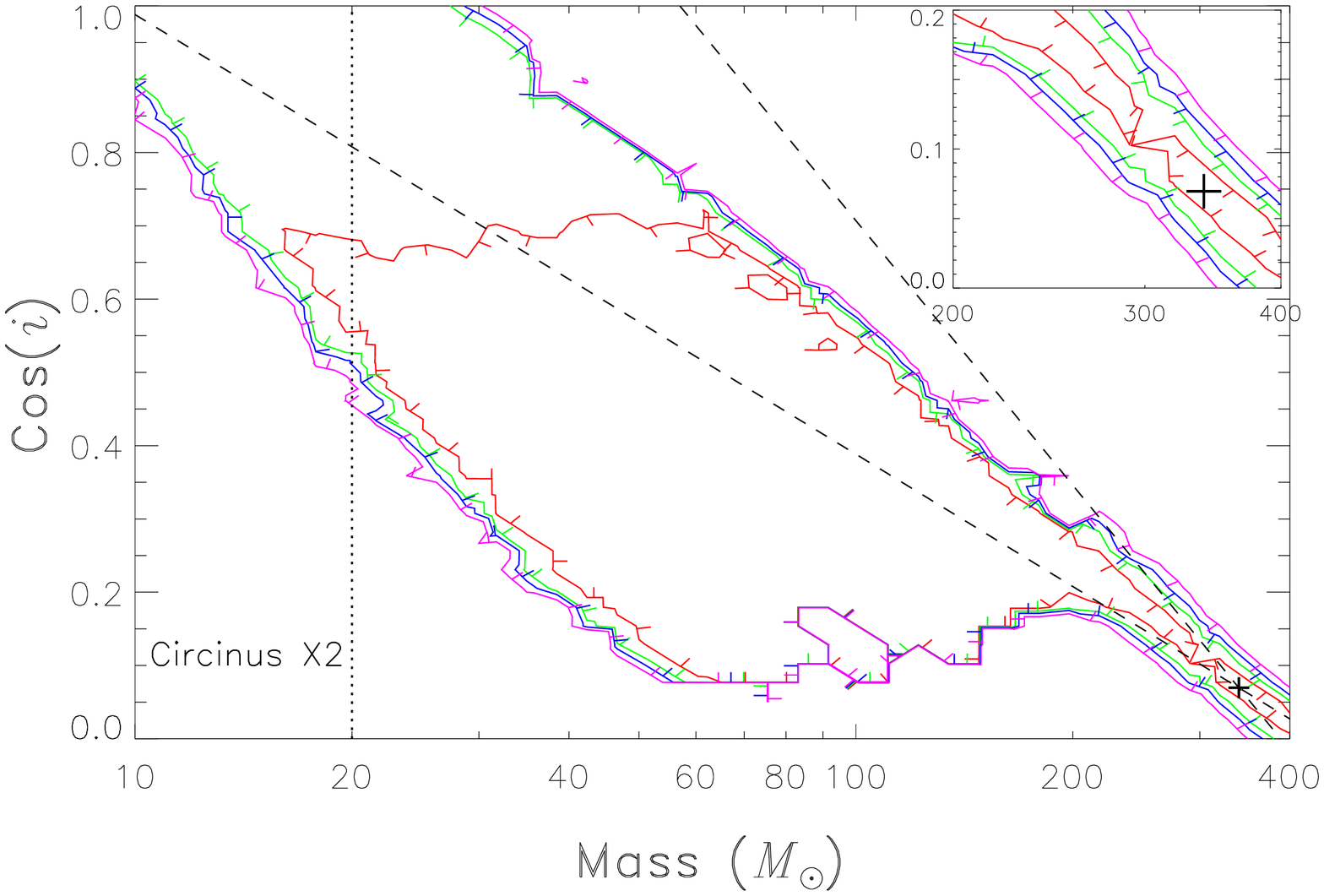}
}
\caption{Confidence level contours in the $M$--$\cos i$ plane. The location of
the least $\chi^2$ is marked by a plus sign.  A pair of dashed lines shows
the extreme Kerr and Schwarzschild degeneracy relations suggested by
\citet{sun1989}. A dotted vertical line at $M =20 M_\sun$ marks the greatest mass consistent
with any of the known Galactic black hole binaries
\citep[][]{mcclintock2004}.  The four contours show values of
$\chi^2$ above the minimum by $\Delta\chi^2$ = 2.30 (red), 6.14 (green),
9.21 (blue), 13.82 (magenta).  These correspond to the $68.3\%$ (1-$\sigma$),
$95.4\%$ (2-$\sigma$), $99.0\%$ and $99.9\%$ confidence
levels, respectively.  Short ticks along each contour show the ``downhill''
direction in $\chi^2$.
\label{fig-mass-incl}}
\end{figure}

We find a very similar degeneracy in our fits. It is illustrated in
the multipanel Figure~\ref{fig-mass-incl}.  These
figures plot the confidence level contours obtained from fitting to the
BHSPEC model as seen in the $M$ -- $\cos i$ plane embedded
in the higher-dimensional $\chi^2$ parameter space.  The value of
$\chi^2$ as a function of $M$ and $\cos i$ is found by
minimizing $\chi^2$ over all the other parameters for those two values
of mass and inclination.  There is indeed a correlation
between the mass and the inclination angle for every object.  In addition,
two dashed lines in the figures show the slope of
the \citet{sun1989} relations, the steeper one for maximal spin, the
shallower one for no black hole spin.  In most cases, as one might
expect, their slopes bracket the slope of the correlation.

This degeneracy explains the large contrast between the BHSPEC and KERRBB
best-fit masses for Circinus~X2. In the former case, the best fit has
$\cos i \simeq$~0 and $M$~$\simeq$~340~M$_\sun$, while in the latter,
$\cos i \simeq$~1 and $M$~$\simeq$~13~M$_\sun$.  Nonetheless, in terms
of the BHSPEC fitting, these two very different answers differ statistically
by only 1 -- $2\sigma$.

With this degeneracy in mind, we now return to evaluating the fitting
results for the other five objects in our sample.  If one assumes a
disk model, three of these --- M81~X1,
M101~X2, and NGC~253~X1 --- require relatively large masses, in all three
cases $>$~25~M$_\sun$.  Smaller masses would lead to very large $\chi^2$,
no matter what other parameters (spin, inclination angle, intervening
column density) are chosen.  At the same time, the fits also suggest that
their masses are not extremely large, generally
$\lesssim$~150 -- 300~M$_\sun$.  Regrettably, however, our confidence
that the formally best constrained of these (M81~X1) has a large mass must be
tempered by the realization that the $L/L_E$ implied by even the highest
acceptable mass for this object is $\simeq 0.85$, a normalized luminosity
so large as to call into question the physical standing of the model
from which it was derived. 

It is also of interest that the three objects with the strongest lower bound
on the inferred mass are exactly the three objects in our sample with the
greatest luminosity, from $\simeq 3$ -- $8 \times 10^{39}$~erg~s$^{-1}$.  Although
M101~X2 and NGC~253~X1 do not have inferred luminosities in Eddington units
as high as that of M81~X1, they are not far below:
$\simeq$~0.3 -- 1 for M101~X2 and $\simeq 0.35$ -- 0.55 for NGC~253~X1.

The remaining two (NGC~253~X3 and X4) have best-fit masses that are several
tens of solar masses, but their data are also consistent with a mass
both considerably lower ($\lesssim$~10~M$_\sun$) and considerably higher
($\gtrsim$~100~M$_\sun$).  Circinus~X2, as we have already discussed, is
also in this category, but with an even wider range of uncertainty.
All three of these have luminosities low enough,
3 -- $10 \times 10^{38}$~erg~s$^{-1}$, that they are at best borderline
ULXs in any case.

Along with the mass and inclination angle, our fits are also sensitive to
the black hole rotation rate.  Viewed in the mass--spin plane, $\chi^2$
rises sharply in the case of the three best-constrained
objects, M~81, M~101, and NGC~253~X-1 when $a/M \lesssim 0.8$--0.9.
There is a slight degeneracy between these two variables, in the
sense that lower spin requires smaller mass, but it is quite weak.
However, we do not regard these results as robust because of a
potential systematic error: our assumption of a Novikov-Thorne
surface brightness profile.

If, as recent work on MHD stresses in accretion disks suggests
\citep{krolik2005}, there are significant stresses throughout the
region of marginally stable (or even unstable) orbits, additional
dissipation there is also likely.  The result would be, for fixed
luminosity, a shift to higher temperature and a smaller effective
radiating area in the emitted spectrum \citep{ak00}.
Employing an estimate of the local dissipation rate derived from simulation
data and a general relativistic ray-tracing code to find what portion of
the radiation reaches infinity, \cite{beckwith2008}
have recently computed the effective ``radiation edge" due to these
effects.  They found that when $a/M \lesssim 0.9$ and the inclination angle
is such that $\cos i \lesssim 0.7$, the characteristic radius for
the radiation seen by distant observers can be displaced inward by
factors of several relative to the prediction of the Novikov-Thorne model.

  The sense of the bias induced by fitting a spectrum on the basis of a
Novikov-Thorne surface brightness profile if the radiation edge really is located
closer to the black hole is to find a mass smaller and/or a spin larger
than the actual one.  A factor of 2 error in the characteristic radius in
gravitational units (i.e., $r/r_g$) translates directly into a factor of
2 in mass because $r_g = GM/c^2$.  In terms of spin, a factor of 2
error would mean that a nominal best-fit of $a/M = 0.9$ should be reinterpreted
as actually indicating $a/M = 0.4$ if the true radiation edge, like the
radiation edge predicted by the Novikov-Thorne model, scales linearly
with the radius of marginal stability.

   Given the several order of magnitude {\it a priori}
uncertainty in our knowledge of the masses of ULXs, a factor of 2 systematic
error in mass is, at this stage, not a major handicap.  Moreover, because we
are especially concerned with determining whether ULX
black hole masses can be as small as those found in Galactic black hole
binaries, the fact that any correction would {\it increase} the inferred
black hole mass means that the Novikov-Thorne estimate is a conservative
one.

   On the other hand, given the much more limited range of
possible black hole rotation rates, a systematic error of order unity
is a major concern in this context.  In addition, the character of the
spectral fitting acts to enhance the magnitude of the possible systematic
error.  Diminishing the spin in a model fit, by moving the
radiation edge outward, also lowers the characteristic temperature
for fixed luminosity and black hole mass.  In order for the spectrum
that results to fit the data, a larger orbital blue shift is required,
forcing the model toward greater inclination angle.  Large inclination
angles are exactly where \cite{beckwith2008} found the largest offsets
between the radiation edge as predicted by the Novikov-Thorne model
and the edge as inferred from simulation data.

\section{Summary}\label{sec:conclusion}
After selecting the 23 known ULXs with the highest signal-to-noise and
least absorbing column density, we chose a subsample of 6 objects in
which the spectrum appeared to be purely that of a thermal disk, with
no hint of any power-law component.  Fitting their spectra with two
different disk models (one based on the multi-color disk approximation,
the other resting on detailed stellar atmosphere calculations), we found
that in all but one case, the results were very similar: the masses
yielding the best fits to the data lie between 20 and 80~M$_\sun$.
In the exception, the two models suggested very different masses,
340~M$_\sun$ in the case of the atmosphere model, 13~M$_\sun$ in
the case of the multi-color disk model.

More significantly, in 3 of the 6, the model fitting (that is, a
search for those parameters for which a disk model can reproduce
the observed spectrum to within the errors) formally excludes
masses similar to those seen in Galactic black holes, $\sim$~10~M$_\sun$.
The models clearly prefer rather larger masses, from several tens of
solar masses to $\sim$~100~M$_\sun$, although in one case the value of
$L/L_E$ preferred by the model-fitting is so large ($\simeq 1$) as to
be physically inconsistent with a disk model.  We also
see formal upper bounds on the mass that are generally
$\simeq$~150 -- 300~M$_\sun$, although systematic error having to
do with the detailed surface brightness profile of relativistic disks
may relax these bounds by a factor of 2 -- 3.  In the other three objects,
the error bars are too broad to permit confident exclusion of either
conventionally small masses or considerably larger ones.

On this basis, we believe that these data provide significant new evidence
that at least some ULXs have masses rather greater --- $\simeq$~30 --
100~M$_\sun$ --- than is found in ordinary Galactic binary black holes.

\acknowledgments{We are grateful to Andy Ptak for extensive instruction
in the proper treatment of X-ray spectral data.  We also thank Shane Davis
and Omer Blaes for many helpful discussions about the calculation of disk
atmospheres.  We especially acknowledge Ivan Hubeny for construction and
maintenance of the stellar atmosphere code TLUSTY and its disk version
TLUSDISK.

This work was partially supported by NASA ATP Grant NAG5-13228 and by
NSF Grant AST-0507455.}

\clearpage

\begin{deluxetable}{llllllll}
\tablecaption{The Disk-Dominated ULX Sources\label{tb-diskULXs}}
\tabletypesize{\scriptsize}
\tablewidth{0pt}
\tablehead{
\colhead{Ob ID} & \colhead{Galaxy} & 
\colhead{D (Mpc)}& \colhead{Object} & \colhead{RA
(h m s)} & \colhead{Dec ($\circ\ \prime\ \prime\prime$)} & \colhead{Tot.
Red. Cts} & \colhead{Red. Ct. Rate ($10^{-2}$ cts~t$^{-1}$)} 
}
\startdata
0200980101 & M81 & 3.6\tablenotemark{1} & x1 & 09 55 32.9 & +69 00 34.8 & 6913, 6524, - &
9.12, 8.00, - \\
0104260101 & M101 & 7.4\tablenotemark{2} &  x2 & 14 03 03.8 & +54 27 37 & 874, 925, 1879 &
3.22, 3.04, 9.55 \\
0152020101 & NGC~253 & 3.73\tablenotemark{3} & x1 & 00 47 32.8 & -25 17 52.6 & 5874, 6316, 16454 &
8.78, 9.24, 28.46 \\
           &        & & x3 & 00 47 35.2 & -25 15 13.8 & 2266, 2360, 5089 &
3.33, 3.41, 8.72 \\
           &        & & x4 & 00 47 23.3 & -25 19 06.5 & 801, 736, 1659 &
1.16, 1.05, 2.82 \\
0111240101 & Circinus & 4.0\tablenotemark{4} & x2 & 14 12 54.2 & -65 22 55.3 & 1391, -, 2703 &
1.51, -, 4.07 \\
\enddata
\tablenotetext{1}{~\citet{freedman1994}.}
\tablenotetext{2}{~\citet{kelson1996,juvce2006}.}
\tablenotetext{3}{~Calculated from the distance modulus given in NED.
Also refer to \citet{karachentsev2003,rekola2005}.}
\tablenotetext{4}{~\citet{freeman1977,iaria2005}.}
\end{deluxetable}

\clearpage

\begin{deluxetable}{lllllllllc}
\tablecaption{Fitting Results\label{tb-fitResults}}
\tabletypesize{\scriptsize}
\tablewidth{0pt}
\tablehead{
\colhead{Galaxy} & \colhead{Object} & \colhead{$n_{H}$\tablenotemark{a}} &
\colhead{$M$/100M$_\sun$} & \colhead{$L/L_e$\tablenotemark{b}   } &
\colhead{Spin\tablenotemark{c}} &
\colhead{Cos $i$\tablenotemark{d}} & \colhead{$\chi^2$/dof} &
\colhead{$L_{obs}/L_{bol}$\tablenotemark{e}}
}
\startdata
M81 & x1 & $1.90^{+0.12}_{-0.11}$ & $0.73^{+0.12}_{-0.06}$ &
$1.00_{-0.15}$ & $0.997_{-0.01}$ & $0.53^{+0.03}_{-0.02}$
& 386.3/397 & 66.2/76.5\\
{\ldots} & {\ldots} & $1.61^{+0.11}_{-0.11}$ & $0.49^{+0.25}_{-0.16}$ &
$0.82^{+0.40}_{-0.28}$ & $0.999_{-0.05}$ & $0.52^{+0.12}_{-0.18}$
& 392.6/397 & -\\

M101 & x2 & $1.18^{+0.18}_{-0.33}$ & $0.57^{+1.21}_{-0.27}$ &
$0.75^{+0.25}_{-0.47}$ & $0.997_{-0.20}$ & $0.59^{+0.15}_{-0.38}$ 
& 156.0/156 & 41.5/46.1\\
{\ldots} & {\ldots} & $0.96^{+0.18}_{-0.15}$ & $0.63^{+1.27}_{-0.60}$ &
$0.38^{+6.11}_{-0.26}$ & $0.999_{-2.0}$ & $0.46^{+0.54}_{-0.37}$
& 157.6/156 & -\\

NGC~253 & x1 & $1.68^{+0.12}_{-0.15}$ & $0.72^{+0.09}_{-0.22}$ &
$0.45^{+0.11}_{-0.10}$ & $0.997_{-0.12}$ & $0.34^{+0.10}_{-0.04}$
& 998.5/963 & 29.2/31.5\\
{\ldots} & {\ldots} & $1.51^{+0.08}_{-0.10}$  & $0.73^{+0.22}_{-0.19}$
& $0.23^{+0.08}_{-0.05}$ & $0.999_{-0.03}$  & $0.28^{+0.09}_{-0.15}$
& 999.3/963 & -\\
       & x3 & $2.93^{+0.15}_{-0.19}$ & $0.50^{+0.31}_{-0.25}$ & 
$0.20^{+0.19}_{-0.07}$ & $0.997_{-0.20}$ & $0.51^{+0.20}_{-0.17}$ 
& 379.6/388 & 10.1/11.0\\
{\ldots}& {\ldots} & $2.79^{+0.16}_{-0.13}$ & $0.63^{+1.13}_{-0.31}$
& $0.09^{+0.09}_{-0.06}$  & $0.999_{-0.20}$ & $0.42^{+0.28}_{-0.33}$
& 380.5/388 & -\\
       & x4 & $0.64^{+0.21}_{-0.21}$ & $0.23^{+0.68}_{-0.13}$ &
$0.14^{+0.10}_{-0.08}$ & $0.997_{-0.34}$ & $0.46^{+0.27}_{-0.42}$
& 143.9/140 & 3.0/3.4\\
{\ldots} & {\ldots} & $0.45^{+0.18}_{-0.27}$ & $0.35^{+0.42}_{-0.30}$
& $0.05^{+0.30}_{-0.03}$ & $0.999^{+0.001}_{-2.0}$ & $0.33^{+0.67}_{-0.25}$
& 145.4/140 & -\\

Circinus & x2 & $5.42^{+0.27}_{-0.38}$ & $3.40^{+12.70}_{-3.30}$ &
$0.04^{+0.60}_{-0.01}$ & $0.996^{+0.001}_{-1.0}$ & $0.07^{+0.68}_{-0.07}$
& 176.4/173 & 7.3/8.6\\
{\ldots} & {\ldots} & $5.45^{+0.46}_{-0.36}$ & $0.13^{+3.28}_{-0.01}$
& $0.34^{+0.03}_{-0.33}$ & $0.501^{+0.498}_{-1.5}$ & $1.00^{+0.00}_{-0.91}$
& 177.6/173 & -\\
\enddata
\tablenotetext{a}{~Column density in units of $10^{21}$ cm$^{-2}$.}
\tablenotetext{b}{~Luminosity normalized to the Eddington luminosity.}
\tablenotetext{c}{~Dimensionless spin parameter of the black hole.}
\tablenotetext{d}{~$i$ - the inclination angle, i.e., the angle between the
disk normal and the line of sight.}
\tablenotetext{e}{~The ratio of the observed luminosity to the model
integrated luminosity. Luminosity is in units of $10^{38}$ erg~s$^{-1}$}
\tablecomments{~1) For each object, the first row shows the fitting results
from BHSPEC and the second row with ``$\ldots$'' shows the results from KERRBB;
2) All uncertainties were computed with $\Delta\chi^2=2.706$, equivalent to
$90\%$ confidence for a single parameter.  3) Where no upper uncertainty
is shown, the data are consistent with all parameters up to the edge of
the parameter space, i.e., for $L/L_E$ the upper bound is unity while
for spin it is 0.997 for BHSPEC and 0.999 for KERRBB.} 
\end{deluxetable}

\clearpage

\end{document}